\documentclass[10pt,sigconf,letterpaper,nonacm,natbib=false]{acmart} 
\AtBeginDocument{%
  }

\setcopyright{acmcopyright}
\copyrightyear{2018}
\acmYear{2018}
\acmDOI{XXXXXXX.XXXXXXX}

\usepackage{graphicx}
\usepackage{xcolor}
\usepackage{caption}
\usepackage{subcaption}
\usepackage{multirow}
\usepackage{hyperref}
\usepackage{booktabs}

\newcommand{\orion}{ORION NT}
\newcommand{\merit}{Merit}
\newcommand{\cu}{CU}
\newcommand{\as}{\textsc{\textbf{AH}}}
\newcommand{\one}{\emph{i}}
\newcommand{\two}{\emph{ii}}
\newcommand{\three}{\emph{iii}}
\newcommand{\four}{\emph{iv}}

\RequirePackage[
  datamodel=acmdatamodel,
  style=acmnumeric,maxbibnames=99
  ]{biblatex}

\begin{document}

\title{Aggressive Internet-Wide Scanners: Network Impact and Longitudinal Characterization} 

\author{Aniket Anand}
\email{{aanand300,dainotti}@gatech.edu}
\author{Alberto Dainotti}
\affiliation{%
  \institution{Georgia Institute of Technology}
  \country{}
}

\author{Jackson Sippe}
\email{jackson.sippe@colorado.edu}
\affiliation{%
  \institution{University of Colorado}
  \country{}
  \postcode{80309}}

\author{Michalis Kallitsis}
\email{mgkallit@merit.edu}
\affiliation{%
  \institution{Merit Network, Inc.}
  \country{}
  \postcode{78229}}

\renewcommand{\shortauthors}{Anand et al.}

\begin{abstract}
Aggressive network scanners, i.e., ones
with immoderate and persistent behaviors,
ubiquitously search the Internet to identify
insecure and publicly accessible hosts. These scanners
generally lie within two main categories; \one) benign research-oriented probers;
\two) nefarious actors that forage for vulnerable victims and host exploitation. 
However, the origins, characteristics and the impact on real networks of these \emph{aggressive scanners}
are not well understood. In this paper, via the vantage point of a large network telescope,
we provide an extensive longitudinal empirical analysis of aggressive IPv4 scanners that spans a period
of almost two years. 
Moreover, we examine their \emph{network impact}
using flow and packet data from two academic ISPs. 
To our surprise,
we discover that a non-negligible 
fraction of packets processed by ISP routers can be attributed
to aggressive scanners. Our work
aims to raise the network community's awareness for these ``heavy hitters'',
especially the miscreant ones, whose invasive and rigorous behavior \one)
makes them more likely to succeed in abusing the hosts they target 
and \two) imposes a network footprint that can be disruptive to critical network
services by incurring consequences akin to 
denial of service attacks. 
\end{abstract}




\maketitle

\section{Introduction}

Intensive and incessant Internet-wide scanning activities have evolved significantly 
over the past several years primarily due to two orthogonal factors:
the development and wide adoption of research tools such as ZMap~\cite{durumeric2013zmap} and Masscan~\cite{masscan}
that have been enabling researchers to examine a plethora of security and networking questions;
and the independent explosion of botnets and malware that target Internet-of-Things (IoT)
applications and hosts (e.g., Mirai and others~\cite{10.1145/2656877.2656893, antonakakis2017understanding,mozi2020,8538636,10.1016/j.diin.2019.01.014}).  
While the utility of \emph{innocuous research scanners} has been indispensable for many applications (e.g., understanding the risk profile and security posture of 
networks and protocols~\cite{10.1145/2810103.2813703,7906943,10.1145/3517745.3561434,10.1145/3471621.3473500},
detecting network outages~\cite{Heidemann2012APA,10.1007/978-3-030-15986-3_14,quan2012detecting}, disclosing and assessing new vulnerabilities~\cite{10.1145/2663716.2663755},
identifying IP space usage and address exhaustion~\cite{10.1145/1851182.1851196,7145335},
studying censorship~\cite{10.1145/3473604.3474562,10.1145/3372297.3417883,7958591} and understanding botnets~\cite{antonakakis2017understanding}
and cybersecurity flaws~\cite{10.1145/2504730.2504755,10.1145/2815675.2815695,10.5555/2362793.2362828,281434,10.1145/2810103.2813707}), their collective impact on
the overall network traffic, their origins, the profile of the applications/ports they target, etc.
are currently not well understood nor have been systematically quantified. 
A similar gap exists in understanding the network impact and characteristics
of \emph{malicious network scanners} (e.g., botnets~\cite{antonakakis2017understanding}
or adversaries that forage for insecure Internet hosts~\cite{Czyz:2014:TPG:2663716.2663717}) 
that are \emph{heavily} probing the Internet. 
In this paper, we attempt to shed some light into the behavior of both families of scanners 
through the lens of \one) a large network telescope and \two) traffic data (i.e., flows and packet streams) 
from several vantage points of a large academic ISP, namely Merit Network,
and a campus university network, i.e., University of Colorado; we collectively 
refer to these probers as \emph{aggressive scanners} (\as, for short, for ``aggressive hitters'') due to their defining
characteristic of exhibiting some sort of ``excessive'' behavior. 


Large \emph{network telescopes} or \emph{Darknets}~\cite{caida_telescope_report,orion} provide a unique perspective
for understanding macroscopic Internet-wide activities, such as scanning~\cite{durumeric2014internet}. Network telescopes
are instrumented to receive and record Internet-wide traffic destined to large swaths of \emph{unused} (but \emph{routed}!)
IP space.
In this paper, we longitudinally study a large network telescope operated by Merit Network,
namely the \emph{ORION Network Telescope}~(\orion)~\cite{orion},
covering about 500,000 contiguous ``dark" (i.e., unused) IPs
for a period spanning 22 months (January 1st, 2021 to October 15th, 2022) to obtain
up-to-date insights into the characteristics of aggressive Internet-wide 
scanners that reach our Darknet. We consider \emph{three separate modalities}
to examine intensive scanning behavior (see Section~\ref{sec:definitions}). 
E.g., following the definition of ``large scans'' from~\cite{durumeric2014internet}, 
we consider hosts that scan $10\%$ or more of the dark IP space to be aggressive. 
Using this definition, we identify 155,010 unique IPs associated with aggressive 
scanning in 2022 across a total of 57,334,643 unique IPs reaching the Darknet. 
They contribute 540 billion packets amounting to $65\%$ of all packets captured in the Darknet for 2022. 

To understand the network impact ascribed to these ``heavy-hitters'' we integrate into our analysis flow data from~\merit,
which serves upwards of one million users. Further,
we examine live streams of packets at one monitoring station at the same ISP and another station at the University of Colorado campus network. 
We join the ISP datasets with the identified hitters to measure the impact 
of the~\as~activities on the network in terms of packet volume.
We found that~\as~packets contribute between 0.1--5.85\%
of the \emph{total ingress/egress packets} processed by core routers on a typical day; this is a non-negligible
fraction. 

Our main contributions include the \one) up-to-date longitudinal profiling of Internet-wide ``aggressive''
scanners and \two) measurable evidence that the aggregate network footprint of these scanners is not as 
inconspicuous as researchers and operators generally assume. This traffic can be disruptive
to network operators; especially traffic originating from origins that never disclose their intents
(as opposed to the seemingly benign ``Acknowledged'' lists~\cite{ackedscanners}
that do reveal the scanning purpose). Scanners of unspecified intent are the vast majority of probers we
categorize as ``aggressive'', and can be associated with botnet propagation and nefarious
reconnaissance (e.g., see~\cite{Czyz:2014:TPG:2663716.2663717}).
We plan to produce and share daily lists of such scanners (using all three definitions) that the network and 
``threat exchange'' communities~\cite{10.1007/978-3-319-45719-2_7,235461}
could subscribe to, hoping that they can be utilized by operators to block and mitigate this disruptive Internet background noise. 

\vspace{-7pt}
\section{Description of Datasets}

\begin{table}[t]
\vspace{-10pt}
\small
\centering
\caption{Description of Datasets.}
\vspace{-10pt}
\label{tab:datasets}
\resizebox{0.89\columnwidth}{!}{%
\begin{tabular}{l|c|c||c|c}
\hline
                         & \textbf{Darknet-1} & \textbf{Darknet-2} & \textbf{Flows-1} & \textbf{Flows-2} \\ \hline
{Packets (Billions)}         & 1,098                         & 833                           & 7,560                   & 770                   \\
{Source IPs (Millions)}      & 123                           & 57                            & 7                        & 2.7                       \\
{Dest. IPs (Millions)} & 0.475                         & 0.475                         & 22                       & 10                      \\
{Total Events (Billions)}    & 26                            & 32                            & -                         & -                        \\\hline
\end{tabular}
}
\vspace{-10pt}
\end{table}

\noindent\textbf{A. Darknet data.}
We analyze data from the \orion~to 
identify and then study the aggressive hitters. 
To study yearly trends, we 
split the Darknet dataset into two parts: \textbf{Darknet-1} (spanning the entire 2021) and \textbf{Darknet-2} (January 1st, 2022--October 15th, 2022). 
See Table~\ref{tab:datasets}.

Central to our analysis of Darknet data is the notion of a \textbf{darknet event}.  
For this study, a darknet event represents a ``logical scan'' such as those defined in~\cite{durumeric2014internet,richter2019scanning}.
Following~\cite{durumeric2014internet}, a \emph{logical scan}
summarizes the scanning activities of a source IP appearing in the Darknet.
TCP-SYN packets, UDP packets, or ICMP ``Echo Request'' packets
are the three \emph{traffic types} we consider as ``scanning packets"~\cite{durumeric2014internet}.
A logical scan represents the activity of a \emph{source IP} 
associated with a particular Darknet~\emph{destination port} 
and~\emph{traffic type}.
For each
darknet event / logical scan we record its \emph{start} and \emph{end} timestamps; an event is considered to have \emph{ended} 
when no packets have been seen in the Darknet from the event's source IP to the event's
targeted destination port and traffic category for more than a ``timeout'' period of around 10 
minutes\footnote{The timeout
or ``expiration'' period is based on ideas from~\cite{caida_telescope_report} (see Section III.E, ``Flow Timeout Problem'')
and the intuition behind it is to avoid splitting ``long scans'' into individual shorter ones. To 
calculate this timeout interval, one needs the Darknet size, an assumed scanning rate and an assumed
duration for the ``long scan''; we used 100pps and 2 days, respectively.}.
For each event, we record \emph{total packets}, number of \emph{unique
Darknet destinations} contacted and metadata~\cite{orion}. 

\noindent\textbf{B. ISP flows.} 
To quantify the scanners' network impact, 
we utilize ISP flows from~\merit. The flows are in Netflow format and collected
with a \emph{packet sampling} rate of 1:1000 at three core \merit~ routers.
The Netflow collectors are configured to only sample \emph{ingress}
and \emph{egress} traffic to/from the ISP.
i.e., internally facing router
interfaces are not included in the flow data.
We employ two datasets:
\textbf{Flows-1} (January 15th, 2022 to January 21st, 2022) and \textbf{Flows-2} 
(October 1st, 2022). 

\noindent\textbf{C. Packet streams.} 
To further validate the network impact results, we also performed
measurements on mirrored packet streams at \merit~and the campus
network at the University of Colorado (to be referred as \cu). \cu~is not associated with 
\merit~(i.e., \merit~does not provide upstream/transit services to \cu~and the IP spaces of both
networks are different), and serves
a population of 100,000 users. These \emph{non-sampled} packet
streams include the majority of ingress/egress traffic observed at a major core router
at \merit~(one of the three routers we have flow data from) and all campus traffic
at~\cu. We examine 72 hours starting on 2022-11-28. During then, at \merit,
the monitoring station processed traffic exceeding 8 Mpps (million packets per second)
and ~$\approx$~80 Gbps. At \cu, we observed peak
rates at 5 Mpps and~$\approx$~40 Gbps. 

\noindent\textbf{D. Acknowledged scanners.}
To obtain insights into the seemingly
benign/research scanners while also partially validating our lists of detected aggressive scanners,
we employ the publicly available list of ``Acknowledged Scanners''~\cite{ackedscanners}. 
The list curator considers a scanning IP as an ``Acknowledged Scanner'' (``ACKed'' scanner, in short)
if the scanners make any efforts to disclose their intentions (e.g., research purposes).
At the moment our analysis was performed, the list \cite{ackedscanners} makes available the source IPs of 36 
unique organizations. 

\noindent\textbf{E. Honeypot data.}
To cross-validate the lists of non-ACKed scanners (i.e., the likely miscreant ones)
and shed light into their behaviors, we employ data from GreyNoise~\cite{greynoise}.
Grey-Noise (GN) operates distributed honeypot sensors at multiple cloud providers 
meticulously placed throughout the world. 
The IPs observed contacting their sensors are tagged by the GN team via an internal process.
An IP is annotated as \emph{benign},\emph{malicious} or \emph{unknown}; more specific \emph{tags}
are also available for some IPs. 
We examined GN data (with 2,962,153 unique IPs) for the whole month
of June 2022.

\noindent\textbf{Ethical considerations.} 
Working with real-world traces requires
ethical and responsible data handling. 
Our measurement infrastructure was designed with careful consideration
and follows best practices imposed by the security/privacy boards and network managers
of the organizations that operate the corresponding instrumentation. For instance,
all of our datasets are passively collected and we never interact or probe any of the identified
IPs present in our datasets. The data were analyzed in a secure manner only by the authors. 
Moreover, we followed the ``code-to-data'' paradigm for analyzing the live packet
streams in which our code was shared with and executed by authorized personnel with access to the mirrored
data. We do not collect nor examine any device MAC addresses or user payload,
and we merely performed packet counting (i.e., total packets originating from~\as)
when examining the packet streams.

Darknet data are generally considered
to pose minimal privacy risks; however, 
we take measures to not expose any identifiable information that
might endanger networks or individuals.
E.g., in the analyses that follow 
we elected to not publicly disclose the actual ASN and
organization names that originate~\as~to
protect the reputation of these networks. 


\vspace{-11pt}
\section{Aggressive Network Scanners}
\label{sec:definitions}


\noindent\textbf{Definition 1: Address Dispersion.}
We classify a source IP appearing in our Darknet as aggressive
whenever it is involved in a darknet event that targets $10\%$ or more dark IPs. 
This definition was also employed in~\cite{durumeric2014internet}
to identify ``large scans''.
We found 2,977,242 scanning events in Darknet-1 and 2,075,485 events in Darknet-2. 
We identified 158,681 distinct IPs satisfying this condition in the Darknet-1 
dataset and 155,010 IPs in 2022. 


\noindent\textbf{Definition 2: Packet Volume.}
The second definition is based on \emph{packet volume}. For each Darknet dataset, we compile
the \emph{Empirical Cumulative Distribution Function (ECDF)} for the number of packets sent per event.
Using the empirical distribution, we calculate the ($1-\alpha$)th-percentile, and declare
a scanner as ``aggressive'' whenever it participates in an event with total packets transmitted crossing the
critical threshold. We utilized $\alpha=0.0001$.

The thresholds that correspond to the top-$0.01\%$ events were found to be 64,810 packets and
23,491 for Darknet-1 and Darknet-2, respectively.
The number of identified aggressive source IPs found from this definition in 2021 was 159,159. 
We noticed that these numbers are very similar to those obtained
using the address dispersion rule; indeed, the \emph{Jaccard similarity 
score\footnote{Given sets $D_i$ and $D_j$, the value $J := |D_i \cap D_j | / |D_i \cup D_j |$ denotes the Jaccard score, where $|\cdot|$ denotes
the set cardinality.}} for the two sets of hitters is found to be 0.8. 
Due to the high similarity among the two populations in the sequel we mostly
focus our attention to scanners identified using the address dispersion definition.

\noindent\textbf{Definition 3: Number of Distinct Destination Ports.}
Our final definition is based on the number of distinct ports that a scanning IP contacts
in the Darknet in a given day. We again source our data to obtain the ECDFs 
for the number of unique ports for both years. 
We use the same $\alpha=0.0001$ to find the critical threshold. 
The ECDFs for Darknet-1 and Darknet-2 differ, indicating a shift
towards more scanned ports (see~Izhikevich~\emph{et al.}~\cite{izhikevich2021lzr} for a possible explanation). 
For Darknet-1, we classified the IPs scanning more than or equal to 6542 ports per day as aggressive,  
whereas for 2022 the threshold is 57,410 ports.  

\vspace{-7pt}
\section{Network Impact}
\label{sec:impact}

Having the lists of \as~ available, we 
now shift focus into understanding the impact that these scanners pose to networks.
First, we utilize flow data from \merit~to measure
the collective packet volume generated by the identified \as~and processed by the ISP's routers
as they transit the network.
We start by individually checking flow data from three core Merit routers.
These routers collectively process more than 50\% of all packets transiting Merit's network.

Table~\ref{tab:flow-routers-impact-def1} showcases the network impact imposed by aggressive
scanners for definition \#1 (we omit results for the second definition since that scanning population is
very similar to the one identified with the first definition; results for definition \#3 show a less pronounced
impact, albeit non-negligible, but we omit them for brevity). We report on the total number of packets observed at
a specific vantage point \emph{originating} from a source IP belonging to an identified \as.
In addition, we also include the \emph{portion of traffic} that these packets amount to with regards to all the packets that a given router processes
for the days examined. The tables highlight a somehow unexpected result: the daily
fraction of aggressive scanners' packet volume lies between $1.1-5.85\%$; this is a relatively high percentage
and indicates that the impact of aggressive scanners on network traffic is not negligible.
To rephrase, we see evidence that, \emph{on average, at least one out of every hundred ingress or egress packets 
 that a router processes is a packet originating from an~\as}.

Table~\ref{tab:flow-routers-impact-def1} illustrates that
the peering arrangements in place at the ISP directly affect the fraction of \as~packets recorded
on a given router. For instance, we
remark that router-1 endures the highest impact with regards to hitters identified with the address dispersion metric;
this can be explained by the fact that definition \#1~\as~frequently originate from Europe and Asia,
as shown in Table~\ref{tab:def1-origin}, 
and router-1's routing policies (e.g., upstream tier-1 peers)
dictate that such traffic would enter \merit~at that point-of-presence. 

We next reflect further on interpreting and validating this surprising result. We note that
the higher percentages occur on weekends, namely when the overall Merit traffic is 
lower. 
We also speculate that \emph{content caching}~\cite{10.1145/3452296.3472928} plays a critical role in ``amplifying'' the effect
of network scanning. \merit~has put in place careful traffic engineering considerations
to have their users benefit from content caches (e.g., videos, etc.) that reside
\emph{within} the ISP. User traffic to/from these content caches does not traverse the 3 border routers we study here
so these packets do not contribute to the calculated ratio. 

\begin{table}[t]
\centering
\caption{\small{Network impact attributed to active~\as~(definition \#1) as seen at the top-3 routers at \merit.} We report the total packets sent by these scanners (in billions) and the percentage of these packets amongst all routed packets.}
\vspace{-10pt}
\label{tab:flow-routers-impact-def1}
\resizebox{\columnwidth}{!}{%
\begin{tabular}{l|c|c|c}
\hline
\textbf{}                        & \textbf{Router-1}                    & \textbf{Router-2}                    & \textbf{Router-3}                    \\ 
\textbf{Date}                            & \textbf{~pkts / pcnt.~} & \textbf{~pkts / pcnt.~} & \textbf{~pkts / pcnt.~} \\\hline
2022-01-15 (Sat)                          & 15.2 (5.82\%)                        & 6.3 (2.84\%)                         & 4.1 (1.9\%)                         \\
2022-01-16 (Sun)                          & 20.4 (5.85\%)                        & 9.0 (3.03\%)                         & 5.4 (1.92\%)                         \\
2022-01-17 (Mon)                          & 19.4 (5.2\%)                        & 8.1 (2.24\%)                         & 5.6 (1.56\%)                          \\
2022-01-18 (Tue)                          & 15.0 (3.62\%)                         & 5.7 (1.51\%)                         & 5.6 (1.2\%)                         \\
2022-01-19 (Wed)                          & 15.1 (3.29\%)                        & 5.5 (1.37\%)                         & 5.6 (1.14\%)                         \\
2022-01-20 (Thu)                        & 14.7 (3.23\%)                         & 5.8 (1.42\%)                         & 5.3 (1.1\%)                         \\
2022-01-21 (Fri)                          & 16.1 (3.66\%)                        & 6.1 (1.56\%)                         & 5.9 (1.38\%)                         \\
2022-10-01 (Sat)                          & 7.9 (2.52\%)                         & 4.9 (1.86\%)                          & 5.6 (2.59\%)                         \\ \hline
\textbf{{Avg} (pkts/pcnt)~} & \textbf{15.5 (4.15\%)}               & \textbf{6.4 (1.98\%)}                & \textbf{5.4 (1.6\%)}    \\\hline  

\end{tabular}
}
\vspace{-10pt}
\end{table}

\begin{figure}[t]
\centering
\begin{subfigure}{0.48\columnwidth}
    \includegraphics[width=0.99\linewidth]{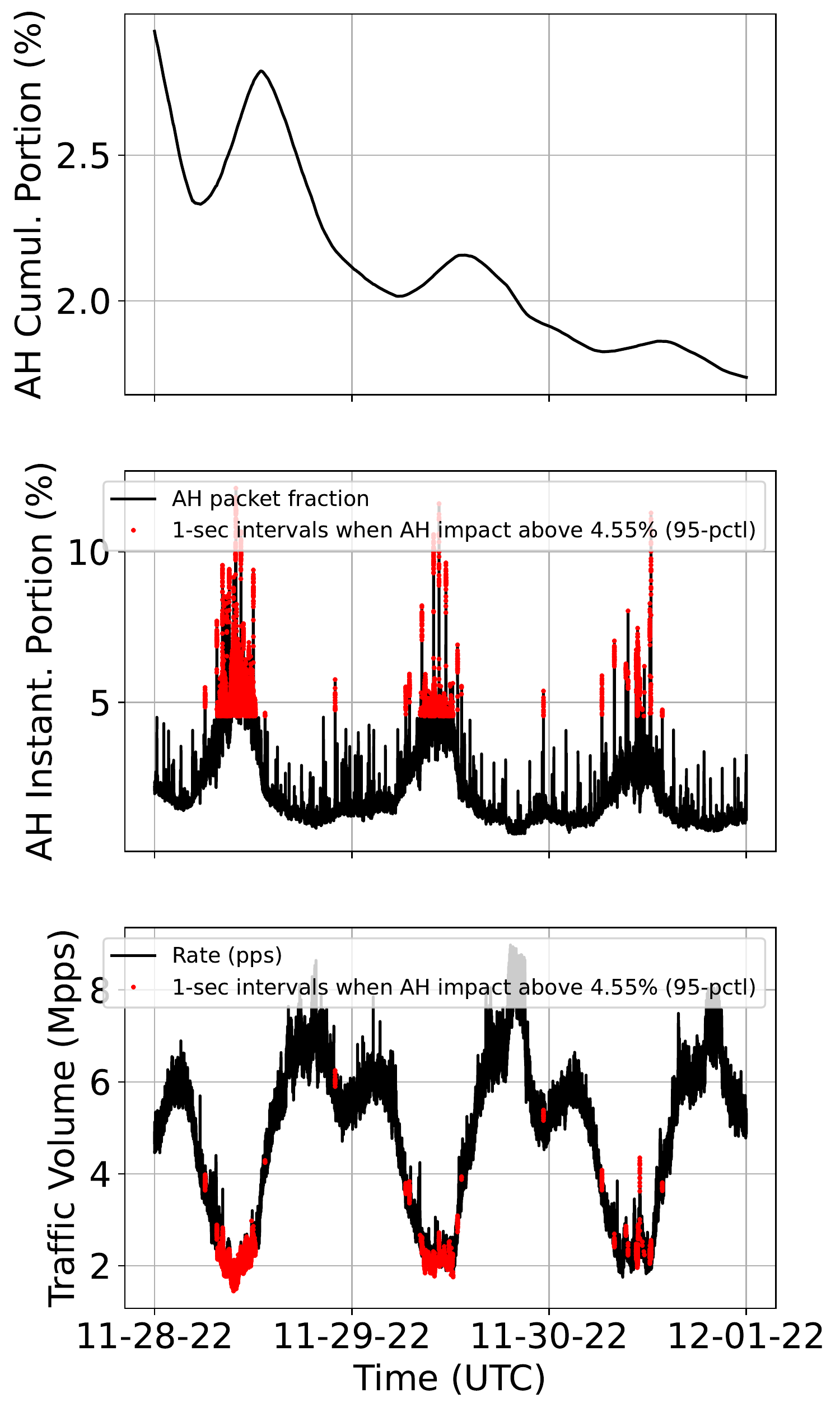}
\end{subfigure}
\hfill
\begin{subfigure}{0.50\columnwidth}
    \includegraphics[width=0.99\linewidth]{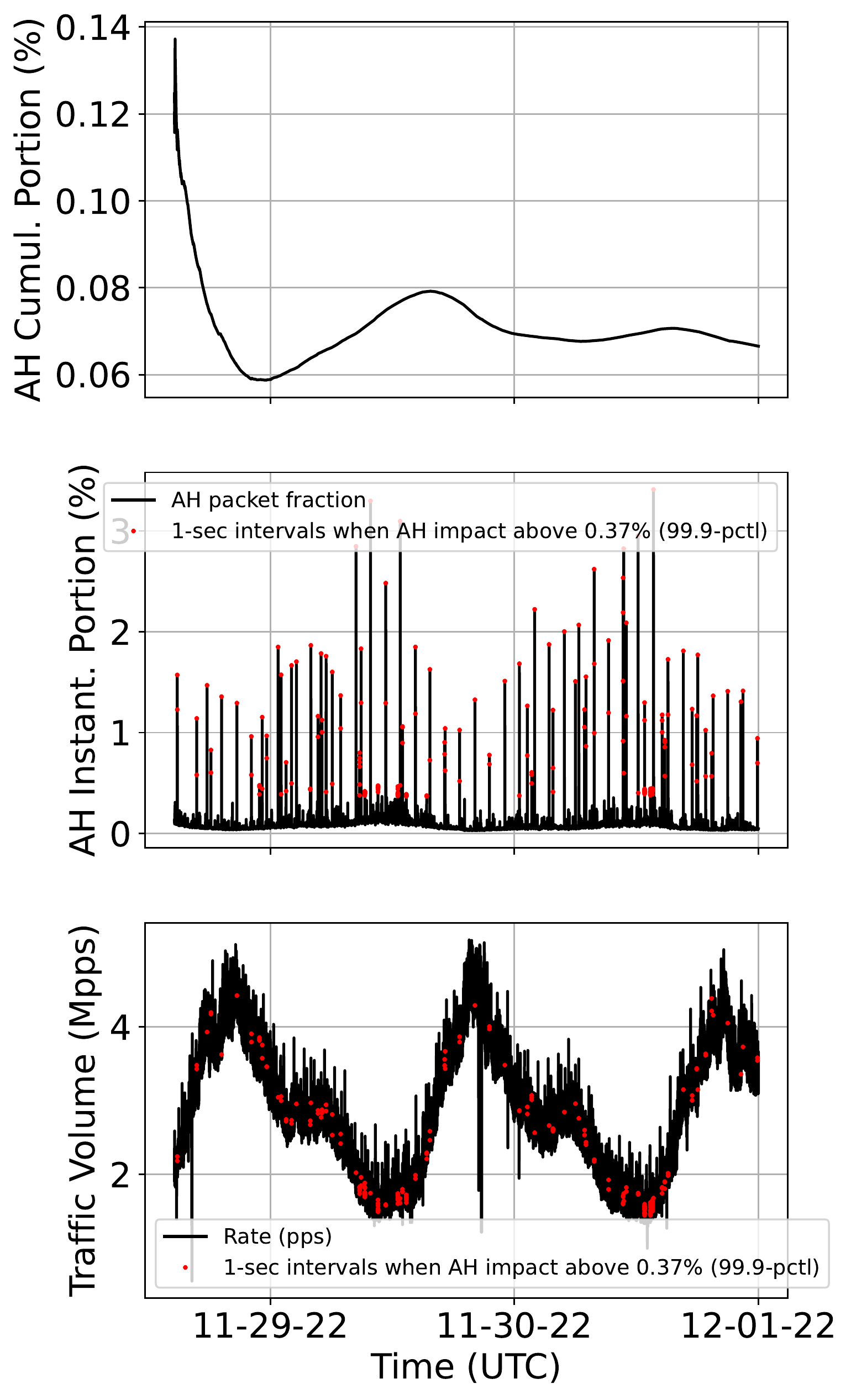}
\end{subfigure}
\vspace{-7pt}
  \caption{\tiny{Network impact (for def. \#1 \as) observed using packet data. 
  Left: \merit~impact. Right:~\cu~impact.
  Top row show the fraction of packets observed at the monitoring station when packets are counted in a \emph{cumulative} manner (i.e.,
  from start of experiment). The center row shows the \emph{instantaneous} impact. Bottom row shows the instantaneous rates; 
  note that on certain occasions (instances highlighted in red), high \as~network impact
  coincides with instances of high overall network traffic rates (in Mpps).}} 
  \label{fig:impact-packets}
  \vspace{-10pt}
\end{figure}

To further validate our results, and to eliminate the possibility that
the high network impact might be due to some bias arising from the sampled flow
data, we next examine the mirrored packet streams at both \merit~and~\cu. 
Figure~\ref{fig:impact-packets} illustrates the results, offering some
interesting findings: 
\one) This non-sampled dataset confirms that the network impact
at \merit~(and router-1, specifically) lies around 2\% (see left panel, 
top row)\footnote{The (cumulative) fraction declines over time since we transition from a weekend day to a
weekday. Further, we performed this 3-day analysis using \as~for Nov. 27th, 2022,
and due to DHCP churn (see~\cite{10.1007/978-3-319-45719-2_7}) some \as~IPs
might have become obsolete by the second and third days of the analysis.};
\two) the network impact at \cu~is also high, but an order of magnitude less than~\merit~(see right panel, top row),
hovering just shy of 0.10\%.
We hypothesized that this could be an artifact of the lack of content caching at~\cu~which
means that the monitoring station at \cu~sees more video-related traffic compared to the~\merit~station.
Indeed, we checked with the network engineers at~\cu~and they verified that no content caching
is present within their network and off-net caching is provided by their upstream ISP;
\three) the instantaneous impact from \as~could even exceed 7\% on certain occasions (middle row panels)
on both networks, reaching even 12\% at~\merit; 
\four) as we observe on the bottom row panels, on several 1-second
intervals (shown in red color) when the \as~impact is high,
overall network traffic could also reach high levels (e.g., exceeding 6 Mpps). This
implies that~\as~are overwhelming the network even during its ``busy'' times,
and consequently network performance might suffer due to potentially incurred packet drops
and network delays. In short, these~\as~collectively exhibit behavior akin to denial-of-service attacks.

\begin{figure}[t]
\centering
\begin{subfigure}{0.48\columnwidth}
    \includegraphics[width=0.99\linewidth]{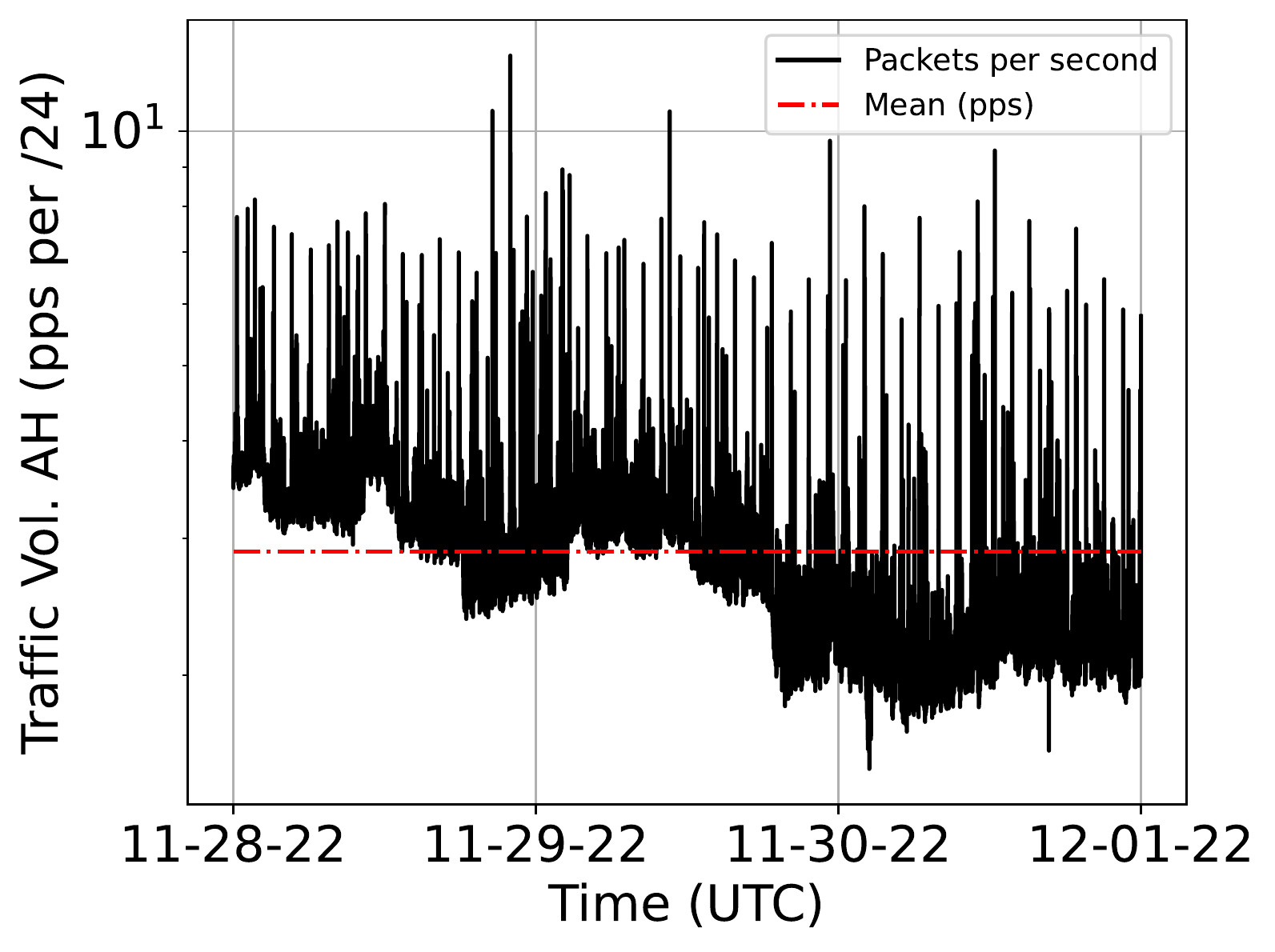}
\end{subfigure}
\hfill
\begin{subfigure}{0.48\columnwidth}
    \includegraphics[width=0.99\linewidth]{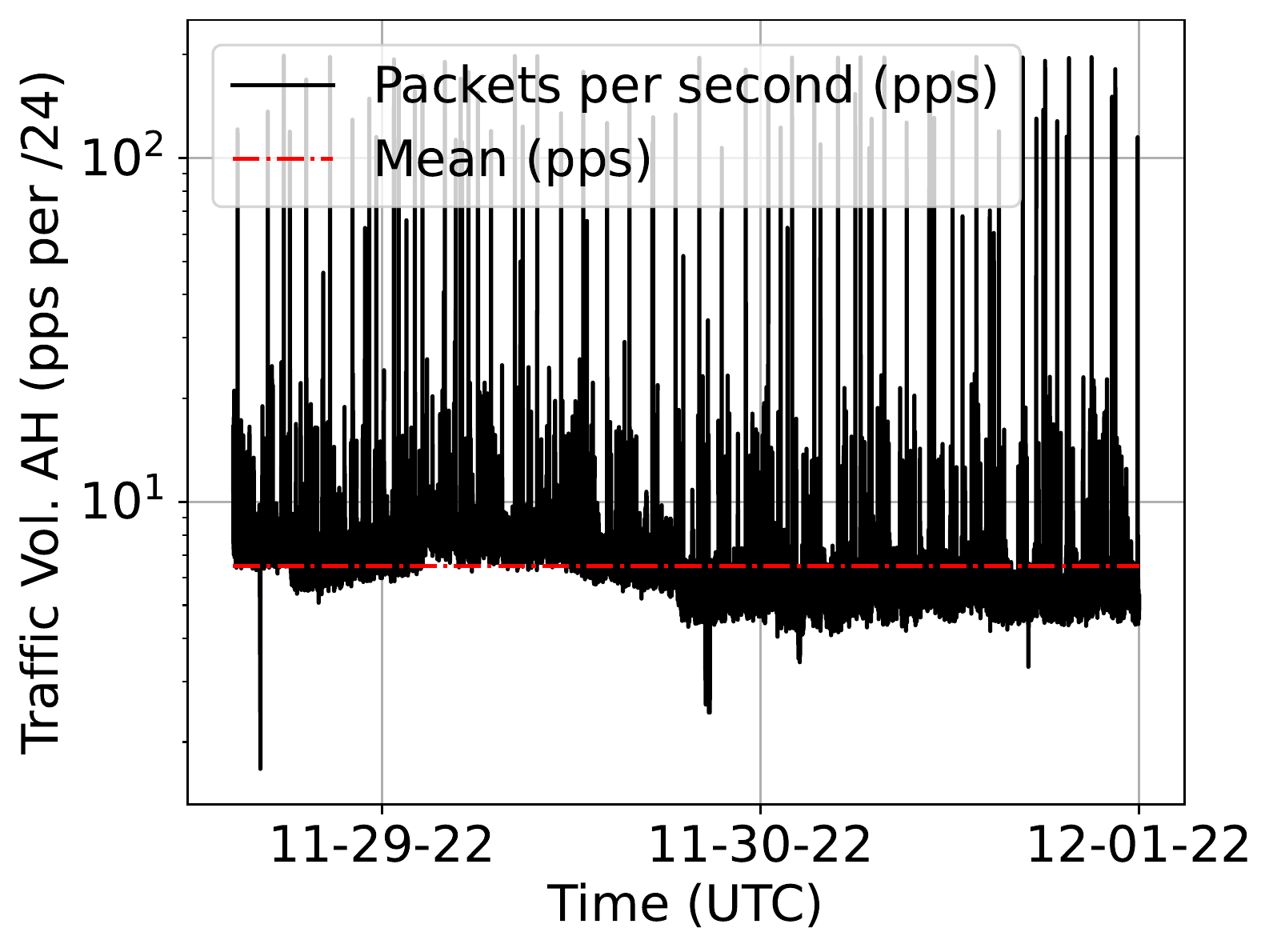}
\end{subfigure}
  \vspace{-15pt}
  \caption{\small{Normalized \as~packet rate by /24s subnets.}} 
  \label{fig:impact-normalized}
\end{figure}

Figure~\ref{fig:impact-normalized} further corroborates the hypothesis that the network impact
difference between \merit~and \cu~can be explained by the presence of content caching (or lack thereof). 
The figure illustrates the instantaneous packet rates ascribed to the identified \as~at \merit~(left)
and \cu~(right) when we normalize by their total number of /24 networks (28561 /24 nets 
for \merit~and 291 for \cu).
As observed, \cu~is in fact more adversely affected by the collective impact of these scanners on a per /24 basis. 

\begin{table}[t]
\centering
\vspace{-10pt}
\caption{\small{Protocols in Darknet (D) and Flow (F)   for 2022-10-01.}}
\vspace{-10pt}
\label{tab:flow-darknet-proto-router1}
\resizebox{0.99\columnwidth}{!}{%
\begin{tabular}{l|c|c|c}
\hline
\textbf{Router-1} & \textbf{Definition \#1}            & \textbf{Definition \#2}            & \textbf{Definition \#3}            \\
\textbf{Protocol} & \textbf{D (\%) / F (\%)} & \textbf{D (\%) / F (\%)} & \textbf{D (\%) / F (\%)} \\\hline
TCP-SYN           & 90.4 / 90.4                        & 88.9 / 89.7                        & 98.2 / 98.7                        \\
UDP               & 9.4 / 8.6                          & 10.8 / 9.2                         & 1.1 / 0.6                          \\
ICMP Ech Rqst     & 0.2 / 0.1                          & 0.4 / 0.2                          & 0 / 0.2             \\\hline              
\end{tabular}}
\vspace{-7pt}
\end{table}


Table~\ref{tab:flow-darknet-proto-router1} allows us to understand
the protocol behavior of these \as, as observed at both the Darknet and Flow data at~\merit.
The table illustrates the protocol distributions with respect to packet volume.
It suggests that the actions of the~\as~are similar across both datasets,
indicating that the high volume of packets we observe originating from them in the flow (and packet)
data is indeed due to scanning and not attributed to other legitimate user behavior originating from the same IPs
that are found to perform scanning.

Table~\ref{tab:acked-pkts} shows
the network impact that scanners that
can be classified as ``Acknowledged''  bear onto the network.
The tabulated data suggest that ``seemingly benign'' scanning activities contribute a relatively
high toll on the routers. The results are for the Flows-2 dataset (October 1st, 2022).

\begin{table}[t]
\centering
\caption{\small{Network impact attributed to ACKed  scanners.
We report total packets sent by ACKed (in billions) and their fraction amongst all ingress/egress packets.
}}
\vspace{-10pt}
\label{tab:acked-pkts}
\resizebox{0.95\columnwidth}{!}{%
\begin{tabular}{c|c|c|c}
\hline
\textbf{} & \textbf{Router-1} & \textbf{Router-2} & \textbf{Router-3} \\ \hline
{Definition \# 1} & 3.17 (1.01\%) & 2.42 (0.92\%) & 5.47 (2.52\%) \\
{Definition \# 2} & 3.35 (1.06\%) & 3.13 (1.19\%) & 5.55 (2.56\%) \\
{Definition \# 3} & 0.5 (0.16\%) & 2.83 (1.08\%) & 0.59 (0.27\%) \\\hline
\end{tabular}
}
\vspace{-10pt}
\end{table}

\vspace{-5pt}
\section{Scanners Characterization}
\label{sec:characterization}

\begin{figure}[t]
    \centering
    \begin{subfigure}[b]{0.5\columnwidth}
        \includegraphics[width=\linewidth]{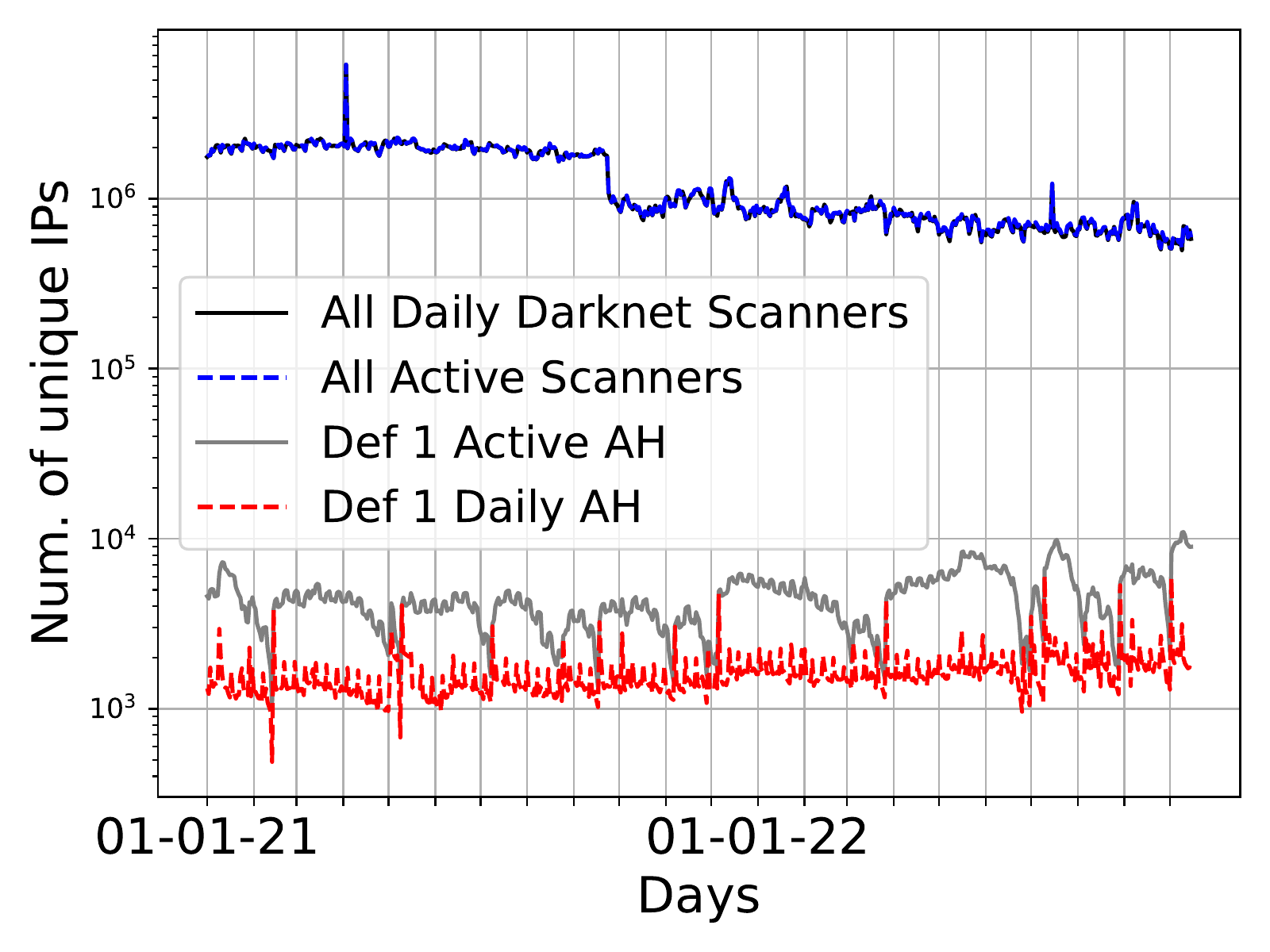}
        \label{fig:timeseries-def1-ipcnt}
    \end{subfigure}%
    \hfill
    \begin{subfigure}[b]{0.5\columnwidth}
        \centering
        \includegraphics[width=\linewidth]{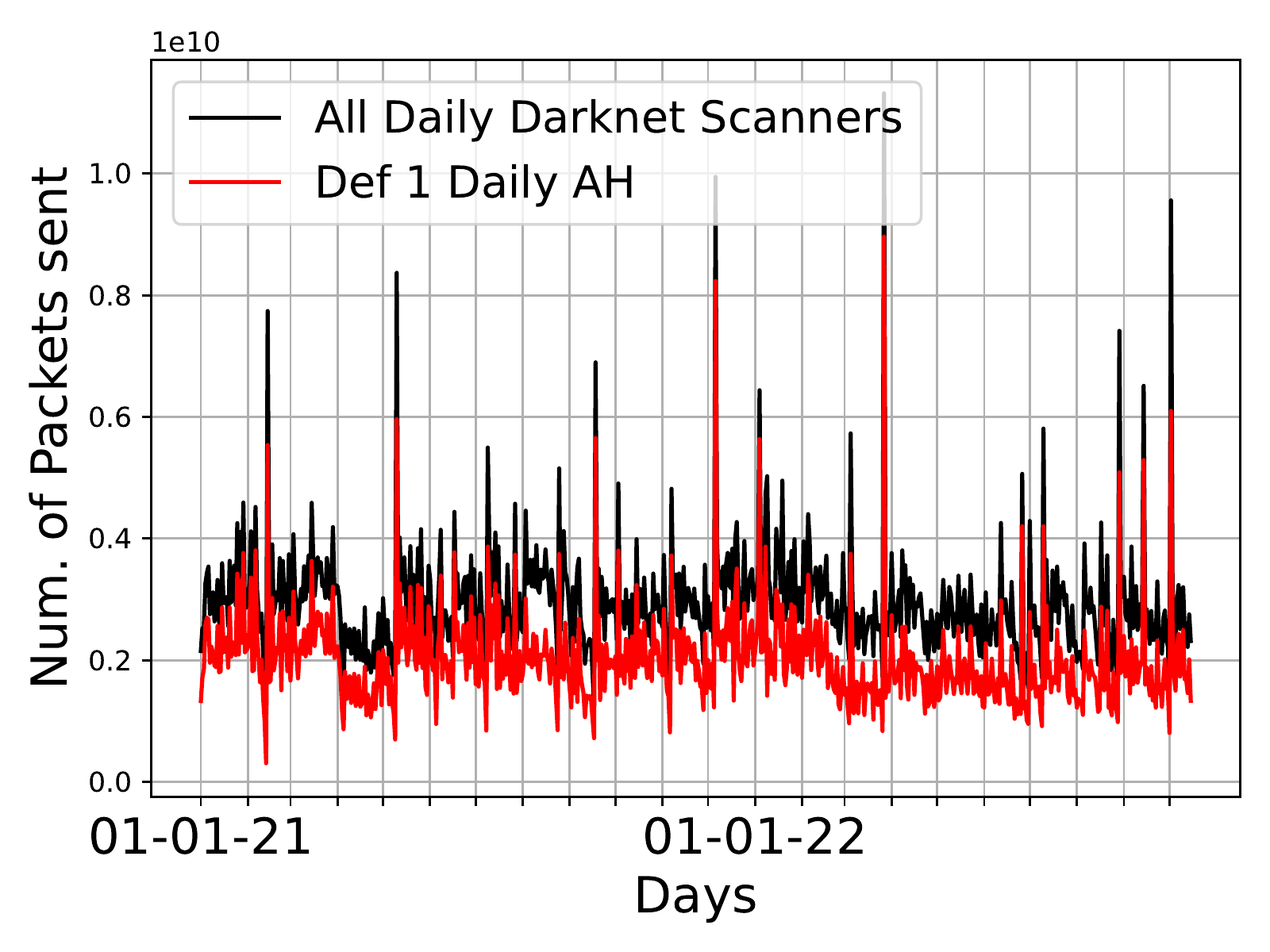}
        \label{fig:timeseries-def1-pkts}
    \end{subfigure}
    \vspace{-40pt}
    \caption{\small{Temporal trends (definition \#1, address dispersion).}}
    \label{fig:timeseries-def1}
\end{figure}

Next, we  longitudinally study the identified scanners and attempt to characterize them (e.g., their origins, top 
ports targeted, etc.).
Figure~\ref{fig:timeseries-def1} shows time-series for definition \#1.
The left panel shows the number of \emph{active}~\as~per day (which includes~\as~that may
have started scanning prior to that day), the number of unique \emph{daily}~\as~(i.e.,
ones that started their scanning efforts during that day), and the number of \emph{all} 
active and daily scanners. The lines for the latter two scanner numbers
seem to coincide because their values are very similar; their average difference is
only 8,471 IPs. 
The right panel shows the number of packets transmitted
by the number of \emph{daily} scanners in a given day, juxtaposed with the aggregate Darknet scanning packets.
Due to the \emph{darknet events} data format, we can only calculate packet statistics for daily scanners.

The plot shows that the number of aggressive scanners increases over time. 
On average, we found 1452 (3876) daily (active) hitters per day in 2021, whereas there are 1779 (5349) daily (active) 
hitters per day in 2022. 
Figure~\ref{fig:timeseries-def1} (right) depicts that the identified hitters contribute the vast majority of packets seen in the Darknet. 
We observe that on average around 0.1\% of scanning IPs appearing in the Darknet and corresponding to~\as~
are responsible for over 63\% of the total packets captured per day in \orion.

Next, we discuss the origins of \as. 
We characterize the type of Autonomous Systems (AS) that originate these scanners,
and the country of origin.
Table~\ref{tab:def1-origin} tabulates the top-10 networks and the countries associated with
definition \#1~\as.
(Numbers in parentheses indicate 
ACKed scanners.)
We also studied the origins of \as~based on the other two definitions;
for space economy, we omit these tables, but we point out that
the origins for the first two definitions are very similar,
echoing the previous observations that scanners from the first two definitions (address
dispersion and packet volume) largely overlap. On the other hand, the origins
for the third group differ, and we even see the presence of research institutions.
Notably, a certain \emph{US-based cloud provider} ranks top in all six definitions/datasets (except once),
indicating strong preference from scanning organizations for its use.

\begin{table}[t]
\caption{\small{Origins of aggressive scanners for definition~\#1.}}
\vspace{-12pt}
\label{tab:def1-origin}
\resizebox{0.99\columnwidth}{!}{%
\begin{tabular}{lccc|lccc}
\hline
 \multicolumn{4}{c|}{\textbf{Darknet-1 (2021)}} & \multicolumn{4}{c}{\textbf{Darknet-2 (2022)}} \\ \hline
 \textbf{AS Type} & \textbf{unique /32s~~} & \textbf{unique /24s} & \textbf{Pkts (B)} & \textbf{AS Type} & \textbf{unique /32s~~} & \textbf{unique /24s} & \textbf{Pkts (B)}\\ \hline
Cloud (US) & 37360 (3799) & 7041 (82) & 65.8 & Cloud (US) & 29933 (3626) & 6601 (67) & 67.2  \\
 Cloud (CN) & 11514 & 7264 & 21.8  & ISP (CN) & 19085 & 10128 & 8.5 \\
 ISP (CN) & 6791 & 5795 & 3.4  & ISP (CN) & 9908 & 7910 & 4.8 \\
 Host. (CN) & 6479 & 4479 & 8.4  & Cloud (CN) & 8777 & 6130 & 19.4 \\
 ISP (TW) & 3753 & 3011 & 1.4  & ISP (KR) & 8228 & 7399 & 3.7 \\
 ISP (CN) & 3601 & 2895 & 4.4  & Host. (CN) & 6657 & 4551 & 11.2 \\
 ISP (RU) & 2708 & 574 & 0.3  & ISP (TW) & 5771 & 4099 & 2.6 \\
 ISP (US) & 2411 & 2166 & 0.2  & Cloud (US) & 3304 & 2955 & 2.5 \\
Cloud (US) & 2364 (250) & 1258 (98) & 4.8  & Cloud (US) & 2891 (54) & 1222 (17) & 3.8 \\
 Cloud (US) & 2248 & 2103 & 2.6  & Cloud (US) & 2244 & 2047 & 1.5 \\ \hline
 Total (\%)    & $79229$ ($50\%$) & 36529 (37\%) & 113.1  (15\%) &  &  95090 (61\%) & 52226 (54\%) & 125.1  (23\%) \\ \hline
\end{tabular}%
}
\vspace{-2pt}
\end{table}

\begin{table}[t]
\vspace{-7pt}
\caption{\small{Validation via ``ACKed Scanners'' lists~\cite{ackedscanners}.}}
\vspace{-12pt}
\label{tab:acked-aggressive-hitter}
\resizebox{0.9\columnwidth}{!}{%
\begin{tabular}{l|cc|cc|cc}
\hline
\textbf{} & \multicolumn{2}{c|}{\textbf{Address Dispersion}} & \multicolumn{2}{c|}{\textbf{Packet Volume}} & \multicolumn{2}{c}{\textbf{Total Ports}} \\ \hline
 & \textbf{2021} & \textbf{2022} & \textbf{2021} & \textbf{2022} & \textbf{2021} & \textbf{2022} \\
IP match & 766 & 766 & 523 & 762 & 317 & 29 \\
Domain matches & 4672 & 4382 & 4334 & 5513 & 71 & 31 \\\hline
Total IPs  & 4706 & 4418 & 4350 & 5549 & 325 & 31 \\
Packets (Billions) & 158.3 & 130.9 & 152.5 & 145.2 & 29.0 & 5.7 \\
Packets (\% all~\as) & 20.4 & 24.1 & 19.9 & 24.3 & 34.0 & 28.1\\ \hline
Total Orgs & 28 & 25 & 27 & 27 & 8 & 4 \\ \hline
\end{tabular}%
}
\vspace{-10pt}
\end{table}

Next, we validate our inferences using the publicly available lists of ``Acknowledged Scanners''~\cite{ackedscanners},
aiming to shed light into organizations that are seemingly benign
and perform aggressive scanning for research purposes.
We consider an identified \as~as an ACKed scanner if 
\one) its IP is within the list of IPs available in~\cite{ackedscanners};
\two) we find a match via reverse DNS checks. I.e., we compiled a list of 48 
``keywords''(see list~\cite{keywords}). 
based on the reverse DNS records of the IPs in~\cite{ackedscanners}.


Table~\ref{tab:acked-aggressive-hitter} summarizes the matching results. 
E.g., we find that 4706 IPs from 27 distinct organizations using definition \#1 and Darknet-1
are indeed \as.
We note that we discovered several IPs (around 7600 in total) belonging to organizations considered as ``ACKed scanners''
that were not included in \cite{ackedscanners}. 
Overall, we identified 7,974 IPs from 29 unique ACKed scanning organizations (out of 36 in~\cite{ackedscanners})
during the full 22-months period across all definitions.



\begin{figure}[t]
    \centering
    \begin{subfigure}[b]{0.5\columnwidth}
        \includegraphics[width=\linewidth]{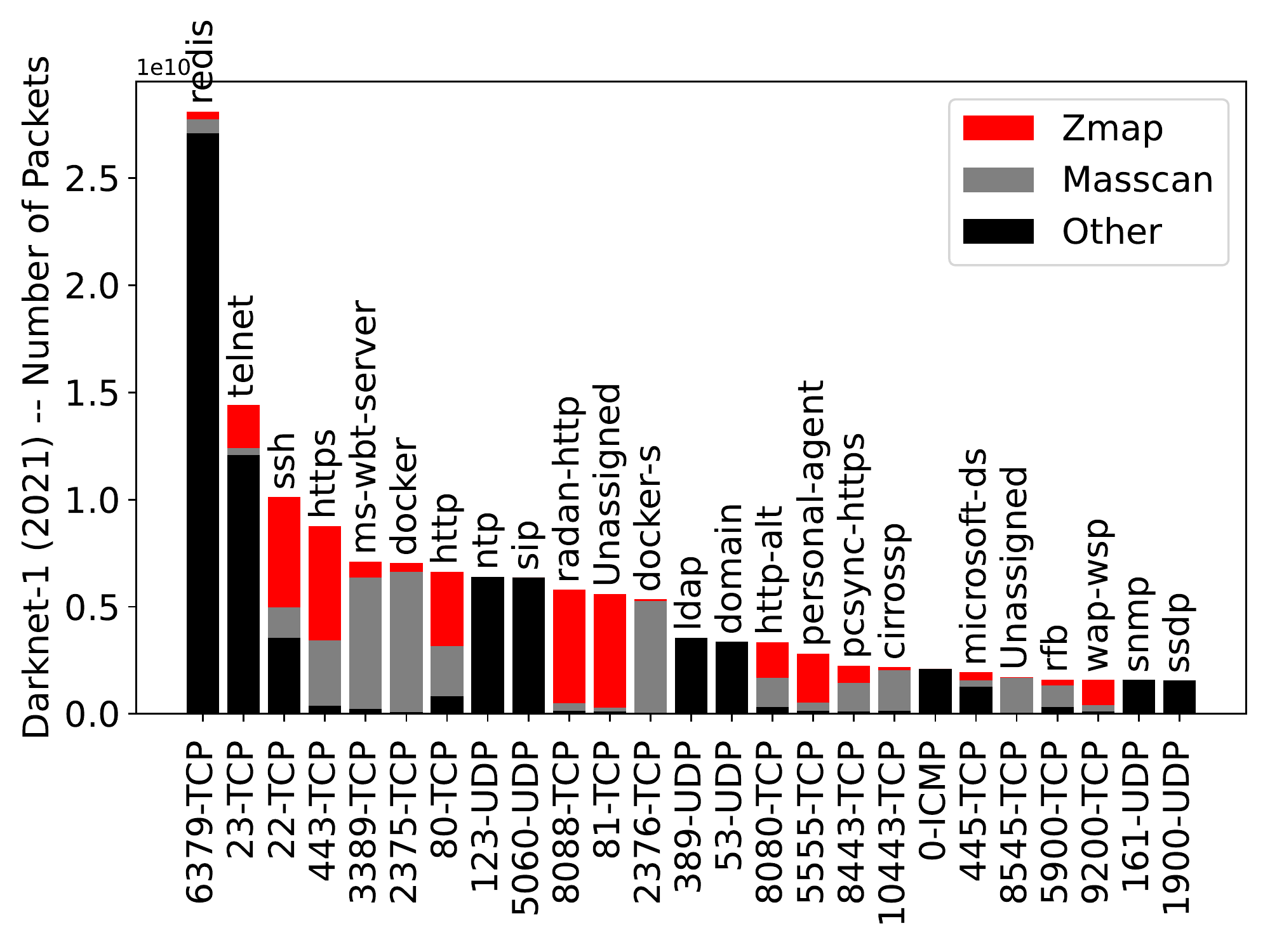}
        \label{fig:port-def1-21}
    \end{subfigure}%
    \hfill
    \begin{subfigure}[b]{0.5\columnwidth}
        \centering
        \includegraphics[width=\linewidth]{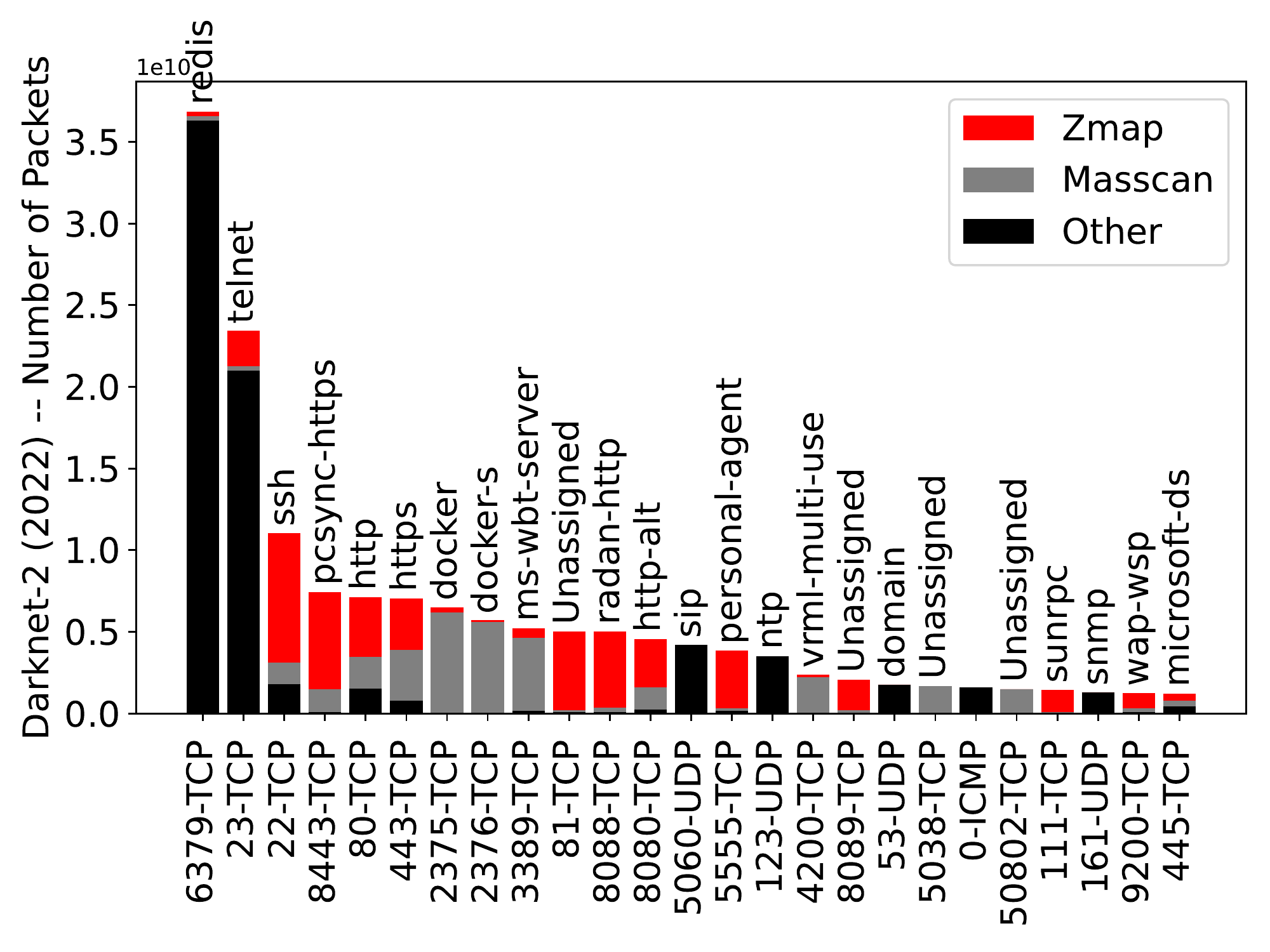}
        \label{fig:port-def1-22}
    \end{subfigure}
    \vspace{-35pt}
    \caption{Top-25 ports targeted by~\as~(definition \#1).}
    \label{fig:port-def1}
    \vspace{-15pt}
\end{figure}

We next characterize the
aggressive hitters in terms of the top
applications they target (with regards to packets received). We also break down the attempts against each port
based on whether the ZMap, Masscan or ``Other'' fingerprints have been observed (see~\cite{durumeric2014internet}
for the ZMap, Masscan fingerprints).
Figure~\ref{fig:port-def1} shows the top ports/protocols for definition \#1. 
We notice that 20 out of top 25 ports are present both in 2021 and 2022, and that~\as~send large number of packets to TCP ports. 
Out of top 25 services which receive the most number of packets in 2021, 
only 4 UDP-based services are targeted.
ICMP (Echo Requests) completes the top-25 set.

Next, we take a moment to compare this behavior
with prior work~\cite{durumeric2014internet}, which also employed Merit's Darknet. 
Figure 2 in~\cite{durumeric2014internet}
shows the same type of~\as~(i.e.,\emph{large scans} targeting more than 10\% of the dark 
IP space) and offers a baseline for comparison. Indeed, \as's~profile
has dramatically changed since the Durumeric~\emph{et al.} 2014 study.
SSH was the top-targeted port by~\as~back then,
but it now ranks 3rd in both 2021 and 2022. The top-ranked aimed
ports currently, according to~\orion, are Redis and Telnet;
neither of them were in the top-5 ports in 2014. This result is somewhat
expected if one considers the rise of IoT applications and the botnet families
that target Telnet services on IoT devices (e.g., see~\cite{10.1016/j.diin.2019.01.014}). 
Further, Redis
vulnerabilities are recently popularly mined for Cryptojacking~\cite{10.1007/978-3-030-58951-6_5}
and other application-level attacks~\cite{9690867}.
Looking at Figure 3
in~\cite{durumeric2014internet}, we also notice that ZMap/Masscan
currently play a prominent role in Internet-wide scanning whereas in 2014
their presence was minimal (as expected, since they were relatively unknown tools then).

Comparing with Richter~\emph{et al.} study~\cite{richter2019scanning},
we do observe some similarities in the top-ranked ports (see Figure 10~\cite{richter2019scanning})
as well as some notable differences. E.g., Telnet was the top-scanned port
in the scanners identified in  Richter~\emph{et al.}~\cite{richter2019scanning},
agreeing with current trends (i.e., Telnet 
is the 2nd most scanned port in our datasets). However, we notice
that Redis/6379 was absent from the rankings of  Richter~\emph{et al.}~\cite{richter2019scanning}.
Interestingly, we also see that TCP/445, one of the most scanned ports in Richter~\emph{et al.}
~\cite{richter2019scanning}, is not preferred by~\as. This agrees with
the results in Durumeric~\emph{et al.}~\cite{durumeric2014internet} where we see TCP/445 mostly
associated with ``small scans" (i.e., scanning
less than 10\% of the Darknet space; see Figure~2,~\cite{durumeric2014internet}).


We also validate our results using lists of scanners obtained from GreyNoise~\cite{greynoise}
in which nefarious aggressive scanners are included. 
Using the month of June 2022 as a basis for comparison, 
we found a significant overlap between the two vantage points;
namely, on average 99.3\% of~\as~identified in our Darknet are also found in GN
on a given day. Since GreyNoise operates a ``distributed'' honeypot
in several regions worldwide, this suggests that most of our identified hitters are not performing localized
scans, but rather engage into macroscopic Internet-wide behaviors.



\vspace{-10pt}
\section{Related Work}
\label{sec:related}



Several notable works have leveraged darknet data to understand IPv4
macroscopic activities; see, e.g.,~\cite{wustrow2010internet, durumeric2014internet, richter2019scanning, 10.1145/3131365.3131383,10.1145/2398776.2398778,10.1007/978-3-030-15986-3_14,10.1007/978-3-319-70139-4_45,
7906943,hiesgen2022spoki,antonakakis2017understanding}.
For instance, network telescopes
have been employed to study malware and botnet outbreaks~\cite{antonakakis2017understanding,10.1007/978-3-319-70139-4_45,10.1145/2398776.2398778}, network outages~\cite{10.1145/2068816.2068818,8784607},
distributed denial of service attacks (DDoS)~\cite{Moore:2006:IID:1132026.1132027,10.1145/3131365.3131383}, trends in Internet-wide scanning~\cite{durumeric2014internet,7906943}, misconfigurations~\cite{wustrow2010internet,10.1145/2504730.2504732}, 
address usage~\cite{benson2015leveraging,10.1145/2567561.2567568}, etc.  Leveraging the large ``aperture'' offered
by large Darknets (i.e., ones that monitor hundreds of thousands or even millions of dark IPs),
one can detect even moderately paced scans within only
a few seconds with very high probability (assuming uniform scans---see~\cite{caida_telescope_report}).

Our study is closest to the works of Durumeric~\emph{et al.}~\cite{durumeric2014internet}
and Richter~\emph{et al.}~\cite{richter2019scanning}. Scanning trends have changed
since these studies were conducted (2014 and 2019, respectively), and we document
some differences in Section~\ref{sec:characterization}. 
To the best of our knowledge, this study is the first that
quantifies the network impact of aggressive Internet-wide scanners. 
We note though that we have not examined IPv6 scanners~\cite{10.1145/2504730.2504732,10.1145/3517745.3561452} nor their impact.
The recent work in~\cite{10.1145/3517745.3561452} studies such scanners through the lens
of a large Content Delivery Network and available firewall logs. We leave analysis
of~\as~IPv6 scanners as future work.

\vspace{-7pt}
\section{Conclusions}
\label{sec:con}

The paper studies a germane sub-population of Internet-wide IPs, namely the~\as~observed
at the~\orion.
The impact on the network of these~\as, as shown in the paper, is surprisingly high.
Thus, understanding their 
behavior is important, with the tangible goal of potentially blocking malicious ones (e.g., the non-ACKed ones)
either at the ``edge'' of an ISP or as they transit the Internet.
An important security implication of these~\as, which are intense and persistent, is that they are
more likely to succeed in finding the vulnerabilities they seek. 
Further, from a network performance perspective, a critical consequence 
is that high packet rates (see Figure~\ref{fig:impact-packets}) from these~\as~could
lead to service degradation akin to ones occurring during DoS attacks.
Thus, raising awareness towards them is important; we plan to
share curated lists of these~\as~with the community on a regular basis.

We offer three concrete methodologies 
on how to identify~\as. 
With the proposed methodologies we aim at obtaining ``quality lists''
of scanners, minimizing false positives due to spoofing or misconfigurations.
Further, succinct \as~lists have practical implications: 
engineers that would consider blocking Internet-wide scanners are 
likely to focus anyways on the top ones
in order to minimize the risk of blocking legitimate traffic due to DHCP IP churn and NAT 
considerations~\cite{10.1007/978-3-319-45719-2_7}.
In fact, as Figure~\ref{fig:gn-analysis} (right, Zipf-like
distribution) in the Appendix shows, even starting by blocking a small amount of~\as,
a large fraction of the problem is ameliorated.


Future plans include further investigating the impact of the aggressive
hitters on more networks beyond the academic ones studied here.
In addition, by examining~\as~observed at additional vantage points (e.g., other large Darknets),
we are aiming to further validate that there is no bias in our existing results.
The fact that we identified~\as~using Merit's ``dark" IP space
and that these~\as~contribute an important traffic portion at a completely different
network (i.e.,~\cu~campus) points towards no selection bias. 
We leave analysis of heavy IPv6 scanners as part of future work, along
with further characterizations of the IPv4~\as~population.




\clearpage
\printbibliography

\newpage
\appendix
\section{Appendix}

\begin{figure}[b]
    \centering
    \begin{subfigure}[b]{0.5\columnwidth}
        \includegraphics[width=\linewidth]{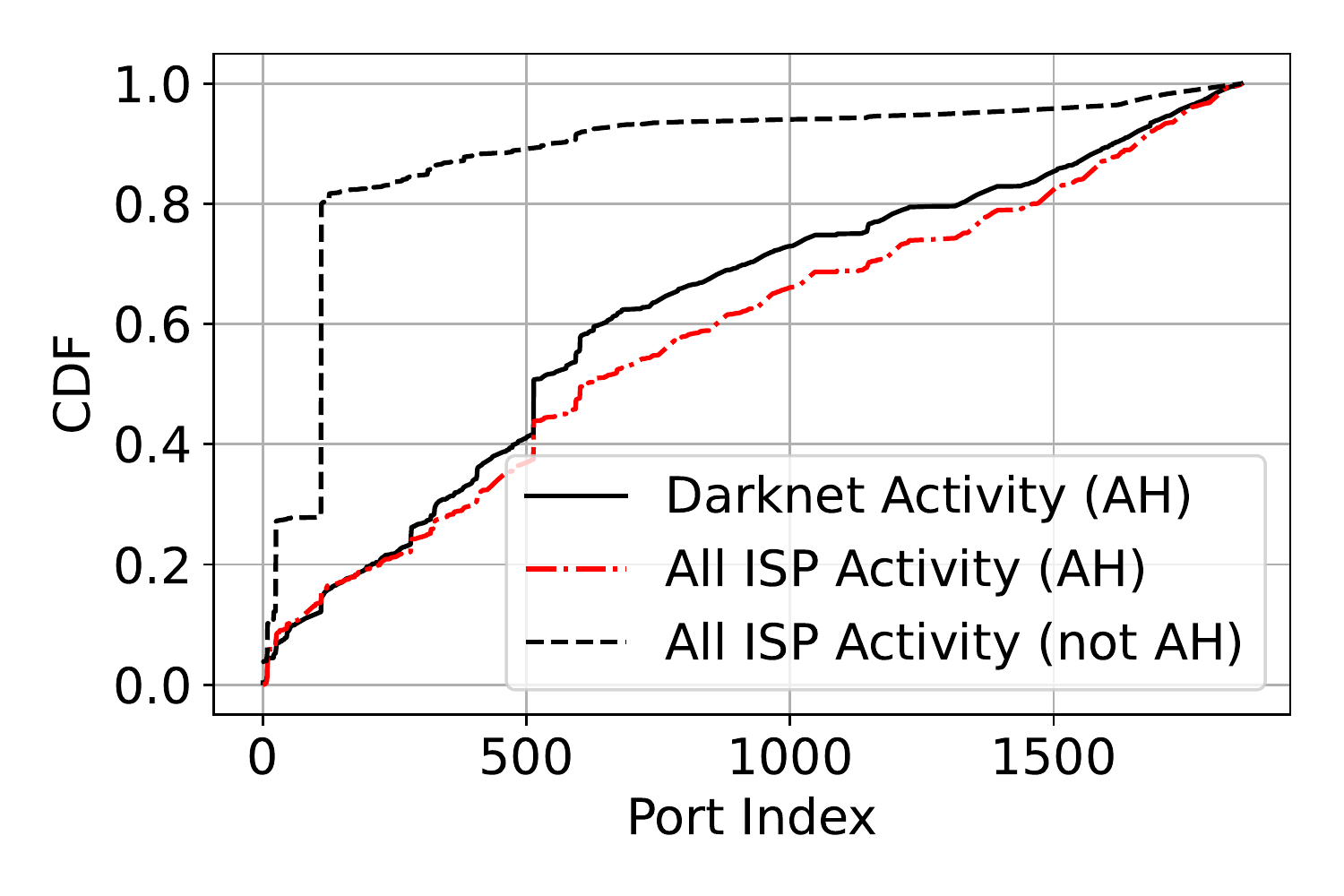}
        \label{fig:port-flow2-def1}
    \end{subfigure}%
    \hfill
    \begin{subfigure}[b]{0.5\columnwidth}
        \centering
        \includegraphics[width=\linewidth]{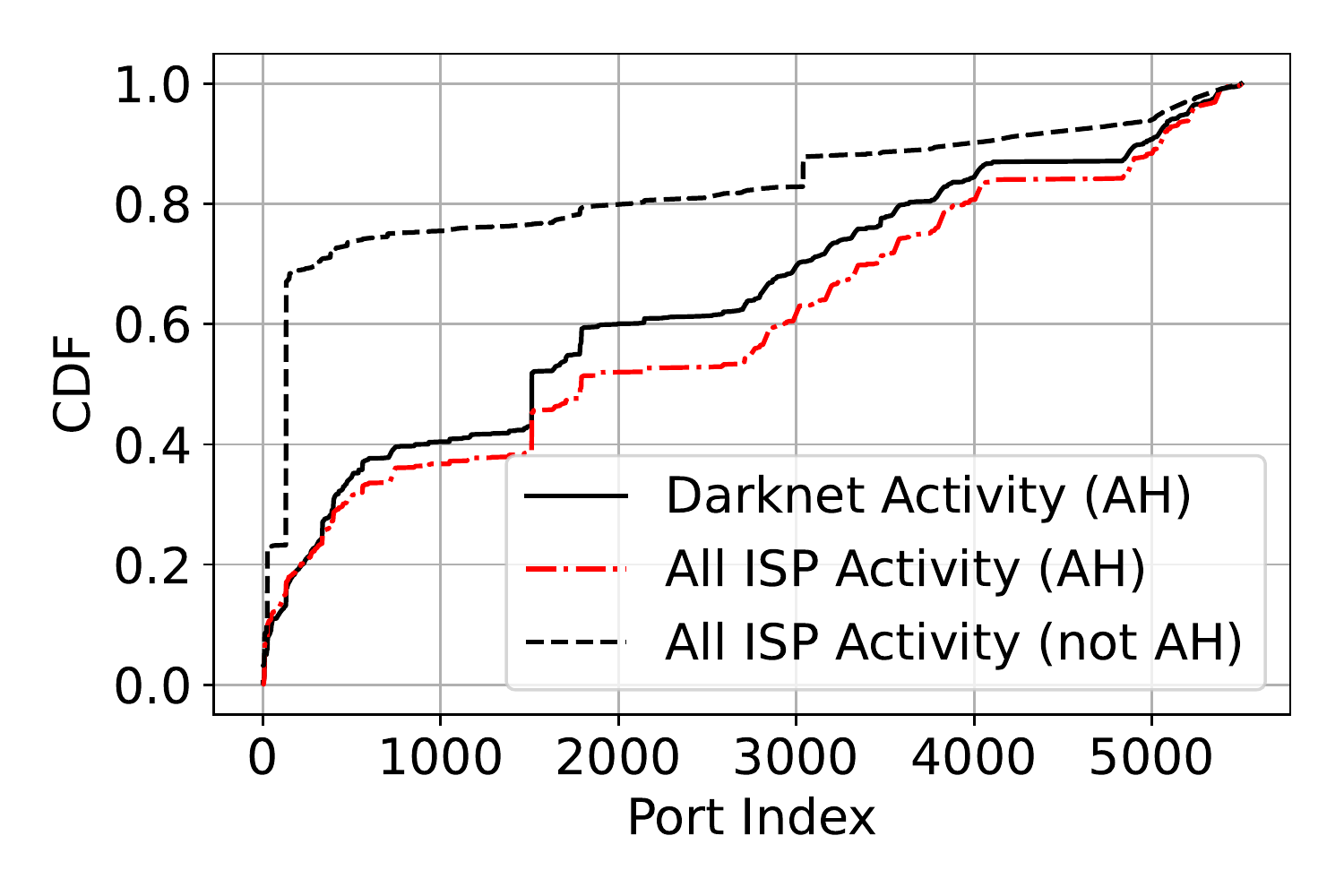}
        \label{fig:port-flow2-def2}
    \end{subfigure}
    \vspace{-25pt}
    \caption{\small{Observed ports in Flow and Darknet (2022-10-01). Left: daily~\as, def. \#1, Right: daily~\as, def. \#2.}}
    \label{fig:port-flow2}
\end{figure}

\begin{figure}[b]
    \centering
    \begin{subfigure}[b]{0.48\columnwidth}
        \includegraphics[width=\linewidth]{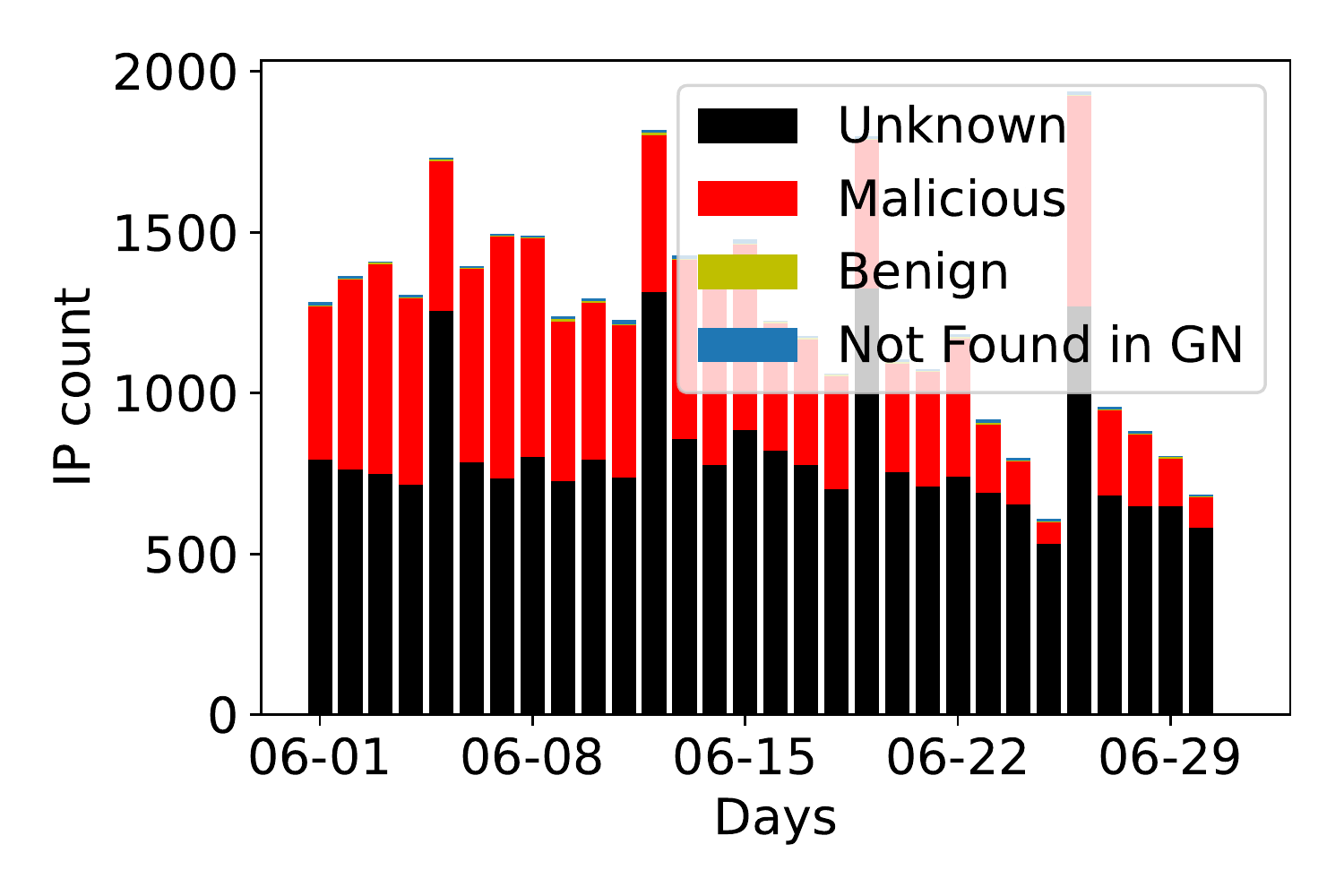}
        \label{fig:gn-class}
    \end{subfigure}%
    \hfill
    \begin{subfigure}[b]{0.5\columnwidth}
        \centering
        \includegraphics[width=\linewidth]{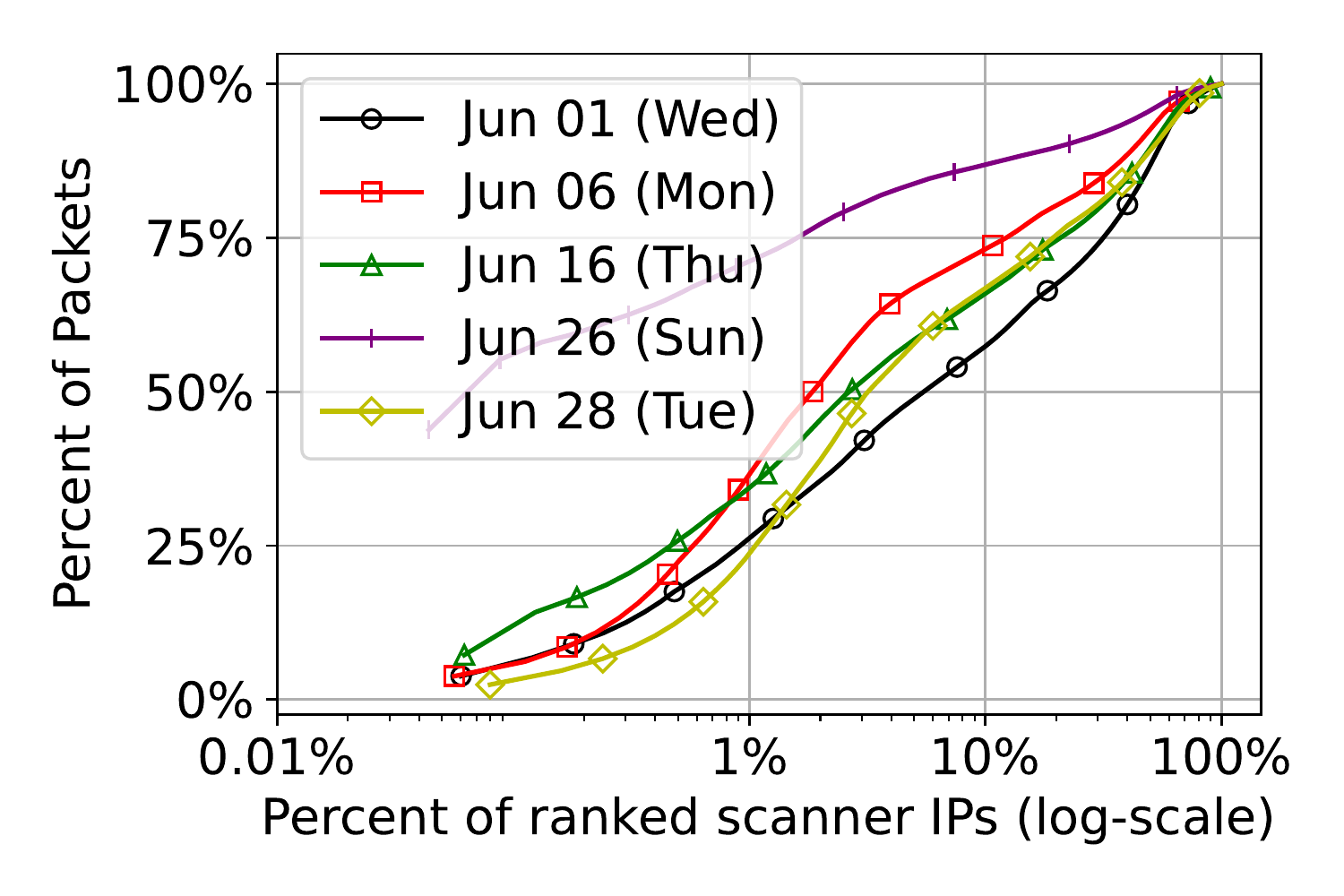}
        \label{fig:zipf-def1}
    \end{subfigure}
    \vspace{-20pt}
    \caption{\small{Left: Breakdown of monthly~\as~within June 2022 based on GN data (def. \#1). 
    Right: Cumulative percentage of all daily~\as~traffic by unique IP
(ranked by packet contribution). The top 1\% of~\as~contribute more than 25\%  on a typical day. Data shown for June 2022.}}
    \label{fig:gn-analysis}
\end{figure}

\noindent\textbf{Supplementary results for Section~\ref{sec:definitions}.}
Table~\ref{tab:def-intersection} 
summarizes~\as~population findings for all definitions and datasets,
and illustrates the intersection of yearly scanners found for all
definitions within the two datasets. 
\newline\noindent\textbf{Supplementary results for Section~\ref{sec:impact}.}
Table~\ref{tab:flow-ipcnt} provides a detailed view of the number of hitters that were identified
using the \orion~and the portion of those that were observed at each vantage point / router.
Figure~\ref{fig:port-flow2}  validates (in accordance with Table~\ref{tab:flow-darknet-proto-router1})
that the actions of the~\as~are similar across both the Darknet and the Flows datasets.
\newline\noindent\textbf{Supplementary results for Section~\ref{sec:characterization}.}
To shed more light into the~\as, we remove the ACKed scanners
and focus on the remaining, presumably malicious, hosts. We leverage GN's threat intelligence
database to obtain insights. Figure~\ref{fig:gn-analysis} (left) depicts the results
in which we consider~\as~identified in June 2022.  
We conclude that \one) a large fraction of the detected~\as~are indeed
malicious, \two) the majority are of \emph{unknown} intentions (thus, merit
further investigation), \three) the benign scanners not removed by our ACKed
scanners filter are very few (hence, the~\cite{ackedscanners} lists
are quite comprehensive) and \four) almost all~\as~identified in~\orion~are
also present in GN, suggesting this population is indeed primarily 
engaged in Internet-wide activities. Table~\ref{tab:greynoise_tags}
zooms-into the top-20 ``tags'' that characterize the set of \as~in~\orion~that are
not ACKed scanners. We observe that a large fraction of these non-ACKed~\as~are
indeed associated with malicious activities (w.g., Mirai-related scanners, worms, etc.).

Figure~\ref{fig:gn-analysis} indicates that even a
small number of~\as~is responsible for high packet volumes.  

\paragraph{Acknowledgements.} We thank the reviewers for all their useful feedback.
This work was partially supported by awards NSF CNS-1823192 and NSF CNS-2120400.

\begin{table}[b]
\caption{\small{Aggressive scanners across all definitions.}}
\label{tab:def-intersection}
\resizebox{\columnwidth}{!}{%
\begin{tabular}{@{}cc|ccc|c|c|c|c@{}}
\toprule
 & \textbf{Darknet-1} & \textbf{D1} & \textbf{D2} & \textbf{D3} & \textbf{D1 $\cap$ D2} & \textbf{D2 $\cap$ D3} & \textbf{D1 $\cap$ D3} & \textbf{D1 $\cap$ D2 $\cap$ D3} \\ \midrule
 & \textbf{IP} & 158681 & 159159 & 3971 & 142012 & 461 & \multicolumn{1}{c|}{426} & 407 \\
 & \textbf{ASN} & 7040 & 6906 & 439 & 6649 & 364 & \multicolumn{1}{c|}{361} & 353 \\
 & \textbf{Org} & 6748 & 6906 & 429 & 6368 & 356 & \multicolumn{1}{c|}{351} & 344 \\
 & \textbf{Country} & 198 & 197 & 80 & 194 & 80 & \multicolumn{1}{c|}{80} & 80 \\ \midrule
 &\textbf{Darknet-2} & \textbf{D1} & \textbf{D2} & \textbf{D3} & \textbf{D1 $\cap$ D2} & \textbf{D2 $\cap$ D3} & \textbf{D1 $\cap$ D3} & \textbf{D1 $\cap$ D2 $\cap$ D3} \\ \midrule
  & \textbf{IP} & 155010 & 295204 & 946 & 155010 & 142 & \multicolumn{1}{c|}{122} & 122 \\
 & \textbf{ASN}  & 5272 & 7837 & 81 & 5272 & 78 & \multicolumn{1}{c|}{74} & 74 \\
 & \textbf{Org} & 5013 & 7470 & 75 & 5013 & 72 & \multicolumn{1}{c|}{68} & 68 \\
 & \textbf{Country} & 183 & 201 & 25 & 183 & 25 & \multicolumn{1}{c|}{25} & 25 \\ \bottomrule
\end{tabular}%
}
\end{table}

\begin{table}[b]
\caption{\small{Number of active \as~ IPs seen on each dataset per Definition (D) and percentage of IPs seen in each router.}}
\label{tab:flow-ipcnt}
\resizebox{\columnwidth}{!}{%
\begin{tabular}{@{}c|c|ccc|ccc|ccc|ccc@{}}
\toprule
 & \textbf{} & \multicolumn{3}{c|}{\textbf{Darknet}} & \multicolumn{3}{c|}{\textbf{Router-1}} & \multicolumn{3}{c|}{\textbf{Router-2}} & \multicolumn{3}{c}{\textbf{Router-3}} \\ \midrule
 & \textbf{} & \multicolumn{3}{c|}{\textbf{\# of \as}} & \multicolumn{3}{c|}{\textbf{Percentage (\%)}} & \multicolumn{3}{c|}{\textbf{Percentage (\%)}} & \multicolumn{3}{c}{\textbf{Percentage (\%)}} \\ \midrule
 & \textbf{} & \textbf{D1} & \textbf{D2} & {\textbf{D3}} & \textbf{D1} & \textbf{D2} & \textbf{D3} & \textbf{D1} & \textbf{D2} & \textbf{D3} & \textbf{D1} & \textbf{D2} & \textbf{D3} \\ \midrule
 \multirow{7}{*}{\textbf{Flow-1}} 
 & 01-15 & 4756 & 7058 & 71 & 97.5\% & 97.3\% &  100\% & 96.0\% & 95.2\% &  100\% & 49.4\% & 48.3\% & 78.9\% \\
 & 01-16 & 5413 & 7794 & 69 & 99.7\% & 99.6\% &  100\% & 98.4\% & 97.8\% &  100\% & 51.8\% & 51.1\% &  82.6\% \\
 & 01-17 & 5466 & 7761 & 69 & 99.9\% & 99.7\% &  100\% & 98.2\% & 97.6\% &  100\% & 51.9\% & 51.4\% &  84.1\% \\
 & 01-18 & 5484 & 7879 & 66 & 99.7\% & 99.4\% &  100\% & 97.8\% & 97.1\% &  100\% & 49.9\% & 49.0\% &  87.9\% \\
 & 01-19 & 4890 & 7361 & 78 & 99.7\% & 99.5\% &  100\% & 98.2\% & 96.4\% &  98.7\% & 52.0\% & 51.4\% & 75.6\% \\
 & 01-20 & 4773 & 7349 & 75 & 99.6\% & 99.4\% &  100\% & 97.7\% & 95.9\% &  100\% & 51.4\% & 50.5\% &  80.\% \\
 & 01-21 & 4662 & 7133 & 92 & 99.6\% & 99.4\% &  100\% & 98.0\% & 96.0\% &  100\% & 51.4\% & 50.3\% &  77.2\% \\ \midrule
\textbf{Flow-2} 
 & 10-01 & 2162 & 3462 & 50 & 94.6\% & 92.1\% &  100\% & 93.7\% & 91.1\% &  100\% & 20.0\% & 19.6\% & 44.0\% \\ \bottomrule
\end{tabular}%
}
\end{table}

\begin{table}[b]
\centering
\caption{GN Tags for non-ACKed \as~(June 2022).}
\label{tab:greynoise_tags}
\resizebox{0.9\columnwidth}{!}{%
        \begin{tabular}{@{}c|c|c@{}}
        \toprule
        \textbf{Rank} & \textbf{GreyNoise Tags} & \textbf{IP Count} \\ \midrule
        \#1 & ZMap Client & 13535 \\
        \#2 & Web Crawler & 11661 \\
        \#3 & Mirai & 8955 \\
        \#4 & Docker Scanner & 4476 \\
        \#5 & Kubernetes Crawler & 4466 \\
        \#6 & SSH Bruteforcer & 1902 \\
        \#7 & TLS/SSL Crawler & 1682 \\
        \#8 & SSH Worm & 1540 \\
        \#9 & Shenzhen TVT Bruteforcer & 1516 \\
        \#10 & Go HTTP Client & 774 \\
        \#11 & Python Requests Client & 765 \\
        \#12 & Telnet Bruteforcer & 720 \\
        \#13 & JAWS Webserver RCE & 693 \\
        \#14 & Ping Scanner & 652 \\
        \#15 & Sipvicious & 624 \\
        \#16 & Looks Like RDP Worm & 509 \\
        \#17 & Carries HTTP Referer & 454 \\
        \#18 & SMBv1 Crawler & 394 \\
        \#19 & Hadoop Yarn Worm & 360 \\
        \#20 & Miniigd UPnP Worm CVE-2014-8361 & 344 \\ \bottomrule
        \end{tabular}
}
\end{table}




\end{document}